\documentstyle[12pt,epsfig]{article}

\newcommand{\be}{\begin{equation}}    
\newcommand{\ee}{\end{equation}}
\newcommand{\beq}{\begin{eqnarray}}
\newcommand{\eeq}{\end{eqnarray}}
\def\lappreq{\! \stackrel{\scriptscriptstyle <}{\scriptscriptstyle \sim}\!}
\def\gappreq{\!\stackrel{\scriptscriptstyle >}{\scriptscriptstyle \sim}\!}
\def\ver{\vskip 12pt}

\def\nn{\nonumber}
\def\op{ \ $ }
\def\cl{$ \ }
\def\ver{\vskip 12pt}
\def\msun{M_\odot}
\def\g{{\bf g}~}
\def\p{{\bf p}~}
\def\s{{\bf s}~}
\def\w{{\bf w}~}
\def\f{{\bf f}~}

\begin{document}

\centerline{\large\bf
{Different approaches to the study of the gravitational radiation}}
\centerline{\large\bf
emitted by astrophysical sources}
\ver
\centerline{\sc Valeria Ferrari}
\centerline{\it Dipartimento di Fisica ``G.Marconi",
 Universit\` a di Roma ``La Sapienza"}
\centerline{\it
and Sezione INFN  ROMA1, p.le A.  Moro
5, I-00185 Roma, Italy}


\begin{abstract}
Stars and black holes are sources of
gravitational radiation in many phases of their life, and
the signals they emit exhibit 
features that are characteristic of the generating process.
Emitted since the beginning of star formation,
these signals also contribute to create a stochastic
background of gravitational waves. 
We shall show how the spectral properties  of this background
can be estimated 
in terms of the energy spectrum of each single event
and of the star formation rate  history,
which is now deducible from  astronomical observations.
We shall further discuss the process of scattering of masses by stars and
black holes, showing that, unlike black holes,
stars emit signals that  carry 
a clear signature of the nature of the source.

\end{abstract}

\section{Introduction}

The extraordinary experimental effort done in recent years to  design and 
construct sophisticated and sensitive gravitational antennas,
allows to  hope that gravitational waves will soon be detected.
When this will happen, will we be  able to
identify the nature of the emitting source,
or to decide whether a  detected stochastic background was  emitted 
during the early stages of the life of our Universe, or later,
by a cosmological population of astrophysical sources?
In order to  answer these questions, waveforms and  energy spectra of 
the signals emitted in different processess have to be 
computed and analysed to extract the  specific
information they provide on   the generating process.
In this paper we shall discuss some astrophysical phenomena
in which this characterization is possible. 

Many phases of the life of a star  are accompanied by the emission of
gravitational waves. For instance, a compact, rotating star emits radiation 
if it has a time varying quadrupole moment due to a triaxial shape or  
to a precession of its angular velocity around the simmetry axes. 
When stellar objects interact,  the
wave  emission can be  associated to the orbital motion, or 
to the excitation of their  proper modes of vibration.
Further on, intense bursts of radiation can be emitted in catastrophic
events such as the gravitational collapse or the coalescence of binary
systems.
All these  processes occurred since  
the beginning of star formation, and therefore they also
contribute to create  a stochastic background of gravitational  
radiation.
To evaluate the spectral properties of the contributions 
of different sources we need a model of the energy spectrum
emitted in each single event, as well as the rate of
events of the selected type occurred  in the recent
or far past, i.e. the  star formation rate  history.
As we shall later see, this important piece of information can be deduced
from recent astronomical observations.

Among the processes that are associated to the emission of gravitational
waves, we shall first consider the  final stages of the life
of a sufficiently massive star which  collapses to form a black hole. 
In this case
numerical simulations show that, unless the collapse is dominated by
very strong pressure gradients or by very slow bounces, the newborn black
hole wildly oscillates and radiates its residual mechanical 
energy in gravitational
waves, at well defined frequencies and damping times that are
associated to its proper modes of vibration: the quasi-normal
modes. We will then show how the energy spectrum emitted in each
event  can be convoluted 
with the observation-based star formation rate history to compute the 
spectral energy-density of the resulting background,
and we will show that it maintains memory of 
the quasi-normal mode signature of each collapse. 
This procedure is applicable to any cosmological
population of astrophysical sources for which a model of  the emitted 
energy spectrum is available.

Another interesting astrophysical process  that is associated
to a well characterized signal is the scattering 
of a mass by  the gravitational field of a massive body, either a
compact star or a black hole. We shall describe how the energy spectra and
waveforms can be computed in the framework of a perturbative approach, and
show that the gravitational signals carry a specific information on the nature
of the emitting source.

\section{The gravitational collapse to a black hole}

In 1939, Oppenheimer \&
Snyder \cite{oppenheimersnyder} studied the collapse of a spherical
cloud of a pressureless gas of particles  ({\it dust}), and showed that
once the collapsing cloud reaches the horizon, it cuts the
communications with the external world and forms a black hole.
In that case, due to the assumed  spherical symmetry, 
no gravitational radiation  emerged. However,
gravitational waves are  emitted if the
collapsing configuration deviates from sphericity,
and this is the basic idea of a perturbative approach  introduced by
C.T. Cunningham, R.H. Price and V. Moncrief \cite{cunningham1,cunningham3}
and subsequently applied by  E. Seidel and T. Moore \cite{seidel1,seidel2}
(for a recent review see \cite{ferraripalomba}).
The perturbative approach consists in the following.
Due to a sudden change in some equilibrium variable,
a sperically symmetric  compact star with an assigned equation
of state collapses.  Typically, the pressure is changed so that
the star is no longer in equilibrium. The collapsing configuration
is obtained  by numerically integrating the Einstein equations 
coupled to the equations of hydrodynamics, and this exact solution
- `exact' in the sense that it is a solution of the fully
non-linear equations -
is used as a background about which linear perturbations are
considered. The perturbed equations,  derived in the framework 
of a gauge invariant formalism \cite{cunningham1,cunningham3}, are then
numerically integrated to compute the energy spectrum and the waveforms.

Alternatively,  the gravitational collapse  can be studied
by directly  integrating the fully  non-linear equations
of gravity, and
at present there is only one fully relativistic numerical simulation 
of collapse to a black hole.  In 1985
R. Stark and T. Piran \cite{starkpiran1} 
numerically integrated   the equations
describing the collapse  of a rigidly rotating,
axisymmetric polytropic star.
Although this simulation is based on some simplifying assumptions, 
it can be considered as a good  model to extract information 
on the collapse of a  massive star to a black hole;
indeed, it has be shown  that
the spectrum of the emitted gravitational energy 
exhibits some distinctive features that are, to some extent,
independent of the initial conditions and of the equation of state
of the collapsing star, and that
depend only on the black hole mass and on its angular momentum.

The dynamics of the collapse proceeds as follows.
If the total angular momentum
\op a=\frac{J}{\frac{GM^2}{c}}\cl is greater than a
critical value \op a_{crit}\cl  
the rotational energy dominates, the star
bounces and no collapse occurs.
Conversely, if \op a < a_{crit}\cl a black hole forms. 
\begin{figure}
\centerline{\psfig{figure=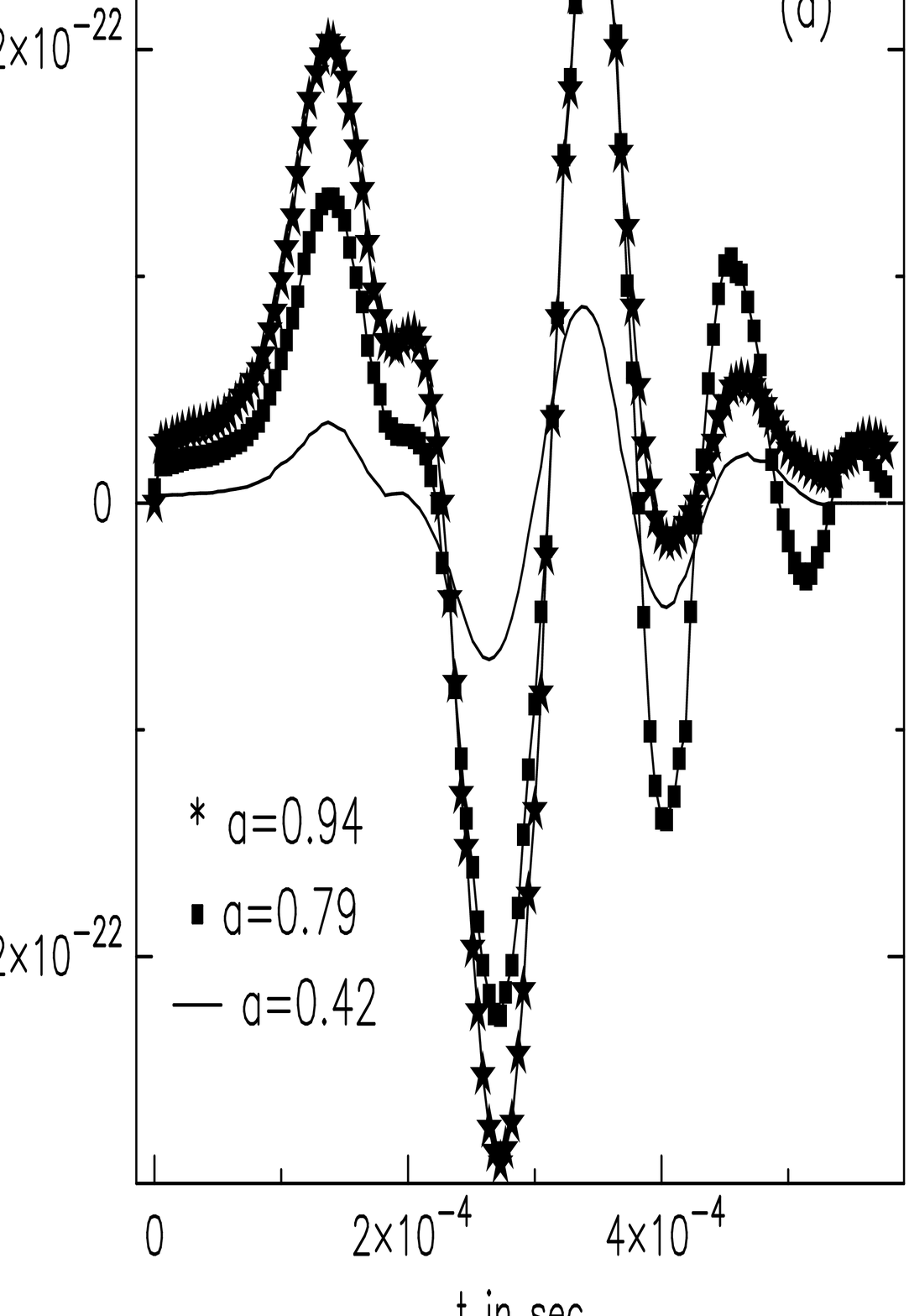,width=8cm,height=6cm}
}
\centerline{}
\centerline{}
\centerline{}
\centerline{}
\centerline{}
\centerline{\psfig{figure=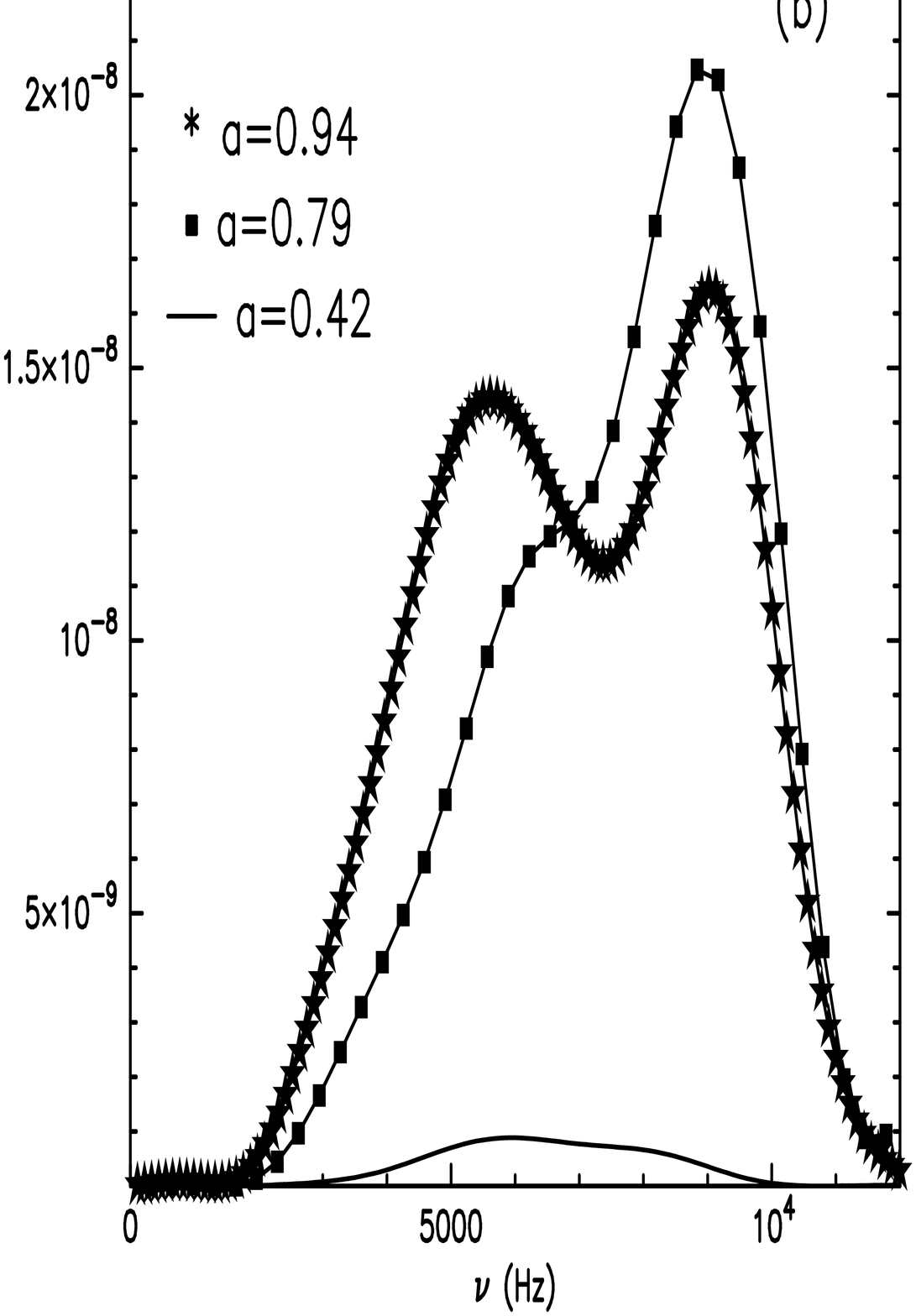,width=11cm,height=6cm}
}
\caption{\it The \op h_+\cl component of the gravitational
radiation (a) and the average energy flux per unit
frequency (b) emitted during the axisymmetric collapse of a
rotating, polytropic star to a black hole, are plotted for
assigned value of the angular momentum
\cite{starkpiran1}. 
These data refer to a collapse ignited by a pressure reduction to a 
fraction \op f_p=0.01$ of the equilibrium central pressure.
}
\label{starkfig}
\end{figure}
In figure 1a we plot the waveform, i.e. the
\op h_+$-component of the metric
tensor of the  gravitational wave emerging at infinity, 
evaluated in the TT-gauge. The \op h_\times$-component is much smaller
and  does not significantly contribute to the emitted radiation.
The function \op f(\nu)\cl plotted in figure 1b is an average energy flux
computed as follows
\be
\label{fdinu}
f(\nu)=
\frac{1}{4\pi r^2}\int_0^{2\pi}d\phi\int_0^{\pi}
\left(\frac{dE_{GW}}{d\nu d\Omega}\right)
\sin\theta d\theta.
\ee
The amplitude of \op f(\nu)\cl scales as the fourth power
on the angular parameter \op a,\cl
and it is proportional to the square of the mass of the collapsing core;
the efficiency of the process is 
$\Delta E_{GW}/M_{core} c^2 \leq 7 \times 10^{-4}$.

Figure 1b shows that if the angular momentum increases
more energy is emitted in the low frequency region.
This is due to the fact that for high values of \op a\cl
the  star becomes flattened into the equatorial plane,
and bounces vertically before collapsing.
However, the more distinctive feature of
these energy spectra  is the peak which occurs
at a frequency that depends on the angular parameter;
for instance \op \nu_{max}\simeq 8.7\cl for \op a=0.79, \cl and
\op \nu_{max}\simeq 9.0~kHz\cl for \op a=0.94.\cl
Moreover, the ringing tails exhibited by the waveforms plotted in figure 1a
are  damped sinusoids at these frequencies.
In order to understand what is the physical phenomenon which associated
to the emission of these peaks, we need to define the quasi-normal modes
of black holes, that emerge
from the theory of black hole perturbations  \cite{MT}.

\subsection{The perturbations of a rotating black hole}

It is known that the radial evolution of the
perturbations of a rotating black hole is described
by a single master equation for a radial wavefunction   \op R_{lm}(r,\omega)$:
\be
\label{teukolski}
\Delta \frac{\partial^2 R_{lm}}{\partial r^2}+
2(s+1)(r-M)\frac{\partial R_{lm}}{\partial r}+V(r,\omega)R_{lm}=0, 
\ee
where
\beq
V(r,\omega)&=&
\frac{1}{\Delta}\Biggl\{ (r^2+a^2)^2\omega^2-4aMrm\omega
+a^2m^2+
2is\Bigl[ am(r-M)\Biggr.\Bigr.\\\nn
&-&\Biggl. \Bigl. M\omega(r^2-a^2)\Bigr] \Biggr\}
+\Biggl[ 2is\omega r-a^2\omega^2-A_{lm}\Biggr]\\\nn
\Delta &=&r^2-2Mr+a^2, 
\eeq
$s\cl is the spin-weight parameter,
\op s=0,\pm 1,\pm 2\cl respectively for scalar, electromagnetic and
gravitational perturbations,
and the sign indicates whether the ingoing  or outgoing radiative part of the
field are considered.  \op A_{lm}\cl is a separation constant.
This wave equation was derived by S. Teukolsky in 1972
\cite{teuka,teukb}.
The quasi-normal modes of a black hole
are defined to be complex frequency solutions of the wave 
equation \ref{teukolski},
that behave as  a pure outgoing wave at infinity,
where radiation has to emerge,
and  as a pure ingoing wave at the horizon,  since
nothing can escape from a black hole horizon.
The eigenfrequency of a mode is complex because the emitted
radiation damps the oscillations, and the
real part is the pulsation frequency, 
whereas the imaginary part is the inverse of the damping time.
Numerical integrations of the wave equations  have shown  that
these frequencies are characteristic of many 
processes involving dynamical perturbations of black holes,
and that an initial perturbation decays, during its very last stages, 
as a superposition of these pure modes.
In table 1 we tabulate the characteristic frequency of the lowest
quasi-normal mode of a  rotating black hole, for \op\ell=2\cl and \op
m=0,\cl   computed by Leaver \cite{leaverqnm}.
The corresponding values in   physical units  can be evaluated by using
the following formulae
\beq
\label{omegatau}
&&\nu_0=\frac{c (M\omega_0)}{2\pi n\cdot M_\odot }
=\frac{32.26}{n}(M\omega_0) kHz,\\
\nn
&&\tau=\frac{n M_\odot}{ (M\omega_i)c}=\frac{n\cdot 0.4937\cdot
10^{-5}}{ (M\omega_i)}~s.
\eeq
where \op M_\odot =1.48\cdot 10^5 cm\cl and
we have  assumed that the black hole mass is
\op M=n M_\odot.\cl
\begin{table}[t]
\caption{The characteristic frequency of the first
quasi-normal mode of a  rotating black hole
is tabulated, for \op\ell=2\cl and \op m=0,\cl
for different values of the angular parameter \op a.$}
\vspace{0.4cm}
\begin{center}
\begin{tabular}{|c|c|}
\hline
$a$ &${M\omega_0+iM\omega_i}$ \\
\hline
0.0    & 0.3737+i0.0890\\\hline
0.2    & 0.3751+i0.0887 \\\hline
0.4    & 0.3797+i0.0878\\\hline
0.6    & 0.3881+i0.0860 \\\hline
0.8    & 0.4019+i0.0822\\\hline
0.9    & 0.4120+i0.0785 \\\hline
0.98   & 0.4223+i0.0735\\\hline
0.9998 & 0.4251+i0.0718 \\ \hline
\end{tabular}
\end{center}
\end{table}
Thus we easily find that for \op\ell=2\cl 
\[
a=0.79\qquad     \nu_{0}=8.6~~ kHz,\qquad\quad
a=0.94\qquad     \nu_{0}=9.0~~ kHz.
\]
If we now go back to the energy
spectra plotted figure 1b, we see that 
they peak at a frequency which is very close to 
\op \nu_{0},\cl i.e. to the 
frequency of the lowest \op m=0\cl quasi-normal mode of the
rotating black hole  which forms as a result of the collapse.
It should be mentioned that the excitation of the quasi-normal modes 
also emerges in simulations of the collapse to a black hole
done in the framework of perturbative approaches.
Thus, the quasi-normal mode peak is a specific signature of the emitted
energy spectrum, since it indicates that the newborn black hole oscillates
in these modes, releasing its mechanical energy in the form of
gravitational waves.

We may ask to what extent is this signature a general feature of the
signals emitted during the collapse of a sufficiently massive star
to a black hole.
A study done by E.Seidel using the
perturbative approach \cite{seidel3}, has shown that if the collapse 
is dominated by extremely high pressures or 
by slow bounces, the infalling material can be considerably slowed
down and the formed black hole may be left with
very little residual energy to radiate away in gravitational waves.
In this case the quasi-normal modes peak may not be as pronounced 
as in figure 1b. 
However, these are quite extreme situations, and it is reasonable to 
presume that the gravitational collapse
will leave behind a newborn black hole with some residual mechanical 
energy, which will be radiated 
in gravitational waves at the frequencies of the black hole
quasi-normal modes.
The efficiency of the gravitational emission
evaluated by Stark and Piran has to be taken as an
indication: it may be smaller, if strong pressures and slow bounces
dominate the collapse, but it may also be higher, if the collapse is
non-axysimmetric.

\section{From a single event to a stochastic background}

\begin{figure}
\begin{center}
\leavevmode
\centerline{
\epsfig{figure=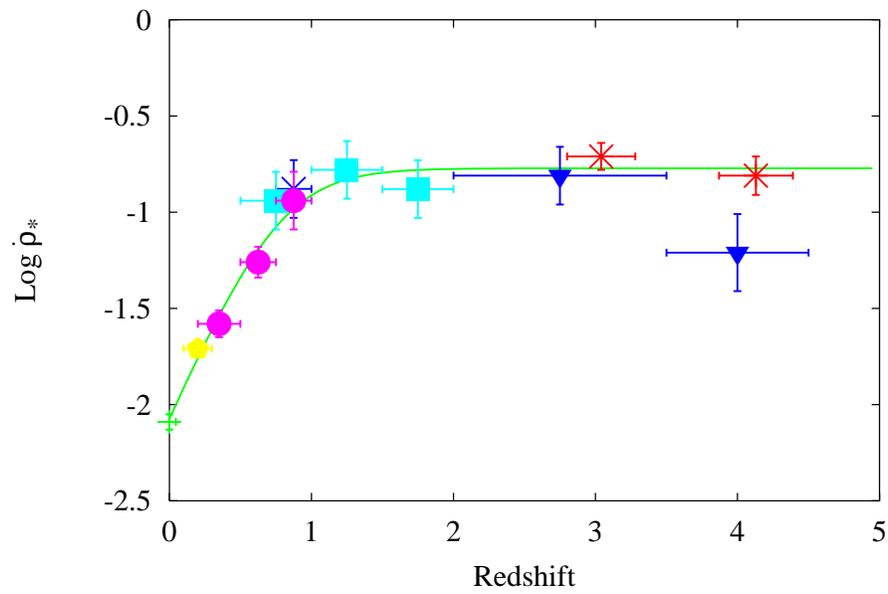,angle=270,width=12cm}}
~~~~~
\vskip 24pt
\caption{The  star formation rate density in units of
$\msun \mbox{yr}^{-1} \mbox{Mpc}^{-3}$ is plotted 
as a function of redshift for a
cosmological background model with  $\Omega_{M}=1$, $\Omega_{\Lambda}=0$,
$H_0=50 \,\mbox{km}\mbox{s}^{-1} \mbox{Mpc}^{-1}$ and a
Salpeter initial mass function.}
\end{center}
\label{sfr}
\end{figure}

The gravitational signals emitted  by newly formed black holes 
throughout the Universe superimpose to form a stochastic background. 
In order to evaluate the spectral properties of this background
and verify whether it keeps memory of the generating process, 
besides the energy spectrum emitted in each single event, we need to know
the rate of events as a function of the cosmological redshift.
This information
can be deduced from the  observative data collected by 
the  Hubble Space Telscope,
Keck and other large telescopes \cite{M96}-\cite{LBG98},
which, together with the completion of several large redshift 
surveys \cite{T98}-\cite{CFRS} have enabled
to derive coherent models for the star formation rate evolution
up to redshifts of $\sim 4-5$.\\
In figure~\ref{sfr} we plot a recently proposed fit  of the data 
obtained at different redshifts, representing 
the star formation rate density \op {\dot{\rho}_{\star}(z)},\cl which
is the mass of  gas which forms stars per unit time and
comoving volume element \cite{M99}, \cite{LBG98}.  
The observative data-points  use
the rest-frame UV-optical luminosity  as an 
indicator of the star formation activity in distant galaxies, 
and it is known that even a relatively small amount of dust can
absorbe the UV light and reradiate it in the
far-IR. This causes an attenuation of the observed UV-luminosity 
(dust extinction) and leads to 
an underestimate of the real star formation activity. To account for this
problem, the data shown in fig.~\ref{sfr} have been corrected  according to
the Calzetti dust extinction law (see \cite{DC}, \cite{LBG98}).
The amount of dust correction to be applied at intermediate-to-high
redshift $z \gappreq 1-2$,  is still relatively uncertain.
However, since the energy emitted by a single source decreases
as the square of the inverse of the luminosity distance,
the gravitational wave backgrounds generated by astrophysical sources
are almost insensitive to the high redshift behaviour
of the star formation rate.
Using the function \op {\dot{\rho}_{\star}(z)}\cl plotted in fig.~\ref{sfr},
the rate of core-collapse supernovae can be computed as follows
\cite{FMS99a}
\begin{equation}
\label{rate}
R(z)=\int_{0}^{z} dz' \, \frac{\dot{\rho}_{\star}(z')}{1+z'} \,
\frac{dV}{dz}
\, \int_{\Delta M} \!\! dM' \, \Phi(M').
\end{equation}
The the factor $(1+z)^{-1}$ takes into account the dilution due to
cosmic
expansion and $\Phi(M)$ is the initial mass function  chosen to be
of Salpeter type, $\Phi(M) \propto M^{-(1+x)}$ with $x=1.7$.\\
The mass range, \op \Delta M,\cl depends on the nature of the considered
source.
Numerical studies on stellar evolution
have shown that  single stars with mass exceeding $~\! 8 \msun$
evolve through all phases of nuclear burning, ending their life
as  core collapse supernovae
(this class includes Type II and Types Ibc supernovae).
While there seems to be a general agreement that progenitors with
masses in the range $8 \msun \lappreq M \lappreq 20 \msun$ leave neutron
star remnants, the value of the minimum progenitor mass which leads to 
a black hole is still uncertain, mainly because of the unknown 
amount of fall back of material during the supernova explosion
\cite{WW95}, \cite{WT96}. 
In the following we shall assume that the lower threshold for
black hole formation is \op  M \gappreq 25~\msun$. 

The rate of events (\ref{rate}) depends also on the cosmological
background through the comoving volume element 
\be
\label{vol}
dV = 4 \pi \left(\frac{c}{H_0}\right) \, r^2 \,
\frac{1}{\sqrt{(1+z)^2\,(1+\Omega_M\,z)-z\,(2+z)\,\Omega_
{\Lambda}}}\, dz,
\ee
where \op r\cl is the comoving distance
\be
r=\frac{c}{H_0 \sqrt{\Omega_\kappa}}\cdot
S\Bigl(\sqrt{\Omega_\kappa} \, \int_0^z [(1+z')^2 \, (1+ \Omega_M \,z') -
\\
z'\,(2+z')\,\Omega_{\Lambda}]^{-1/2} \, dz'\Bigr),\nonumber
\ee
and the function \op S \cl is given by \cite{kim}
\beq
\Omega_M +\Omega_{\Lambda}>1
&\quad S(x)=\sin(x)
&\quad \Omega_\kappa=1-\Omega_M +\Omega_{\Lambda} \\
\nn
\Omega_M +\Omega_{\Lambda}<1
&\quad S(x)=\sinh(x)
&\quad\Omega_\kappa = \Omega_M +\Omega_{\Lambda}-1 ,\\
\nn
\Omega_M +\Omega_{\Lambda} =1
&\quad S(x)=x
&\quad \Omega_\kappa=1 .
\eeq
If we  consider the following cosmological backgrounds
\footnote{
We shall assume
\[
h=\frac{H_0}{100}~km/s/Mpc,
\]
where \op H_0\cl is the Hubble constant.
}
\beq
 A)& \Omega_{M}=1   & \Omega_{\Lambda}=0    \qquad h=0.5 \\\nn
 B)& \Omega_{M}=0.3 & \Omega_{\Lambda}=0.7  \qquad h=0.6 \\\nn
 C)& \Omega_{M}=0.4 & \Omega_{\Lambda}=0   \qquad h=0.6
\eeq
the total rate  of core-collapse supernovae 
leading to a black hole or to a neutron star ranges, respectively, between
\be
\label{ratetot}
R_{BH} ~\sim 3.3 - 4.7 ~events/s, \qquad R_{NS}~\sim  ~ 13.6 - 19.3~ events/s.
\ee
The main difference between the three cosmologies is introduced by the
geometrical effect of the comoving volume, and it is significant at
$z \gappreq 1-2$. This implies that the gravitational backgrounds, which
are mainly contributed by sources at $z \lappreq 1-2$,
are almost insensitive to the cosmological parameters.\\
It is now possible to compute the spectral energy density of the gravitational
wave background produced by an assigned population of sources
\be
\label{spec}
\frac{dE\left(\nu\right)}{dt dS d\nu}=
\int^{0}_{\infty}\int_{\Delta M}  ~ \Biggl[
f(\nu)\cdot d R(z,M)
\Biggr],
\ee
where \op d R(z,M)\cl is the differential rate, and \op f(\nu)\cl
is the average energy flux per unit frequency,
chosen as a model for the class of sources under consideration,
emitted by a source located at a luminosity distance
\op d_L(z) = (1+z)~r,\cl
\be
\label{fdinulum}
f(\nu)=
\frac{1}{4\pi d_L(z)^2} 
\int_0^{2\pi}d\phi\int_0^{\pi}
\left(\frac{dE_{GW}}{d\nu d\Omega}\right)
\sin\theta d\theta.
\ee
For instance, for core collapses to a black hole we shall use as
a model of \op f(\nu)\cl
the energy spectrum (\ref{fdinu}) shown in figure 1b,
rescaled with the luminosity distance and suitably redshifted.

\begin{figure}
\begin{center}
\leavevmode
\centerline{\epsfig{figure=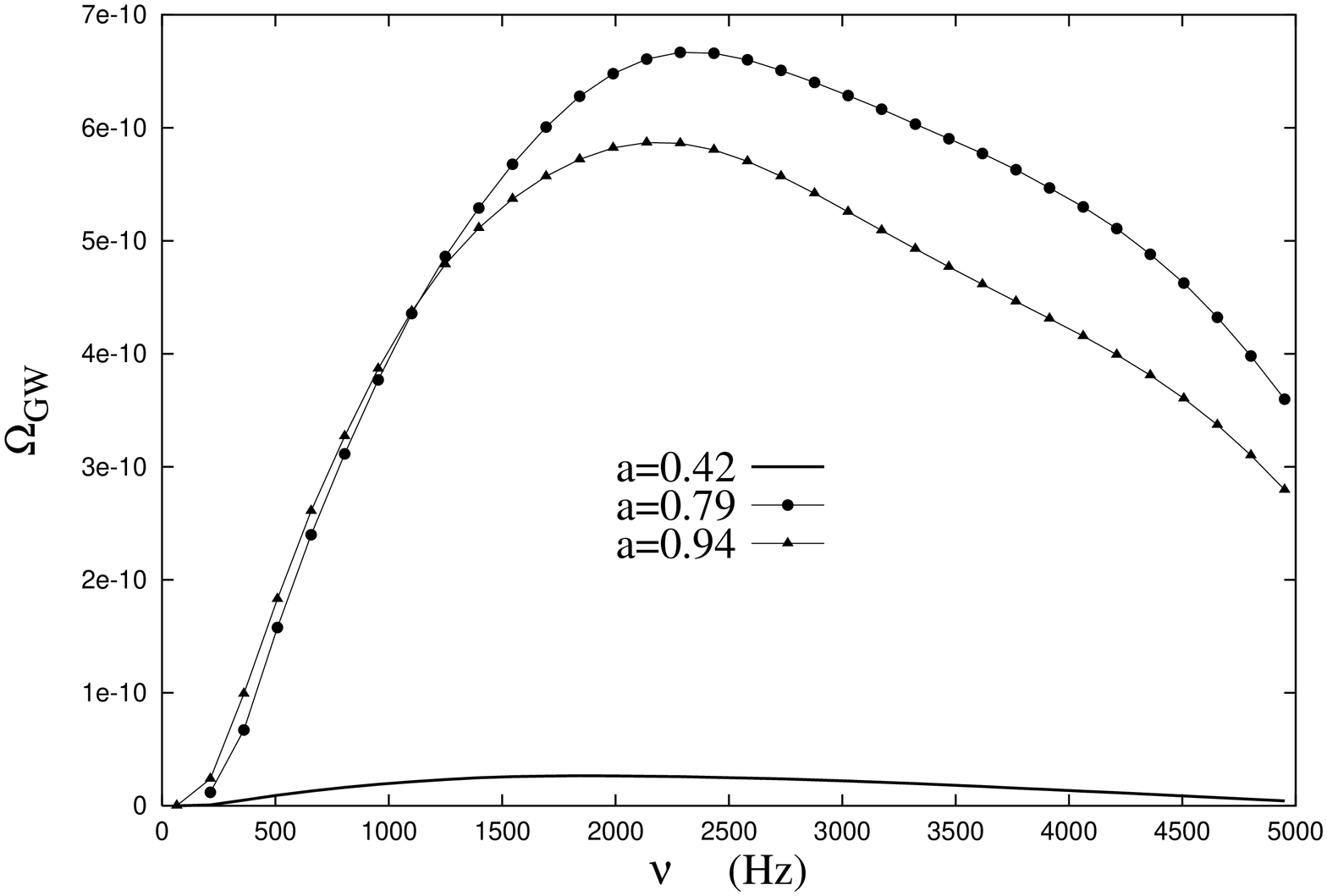,angle=270,width=8cm}
\epsfig{figure=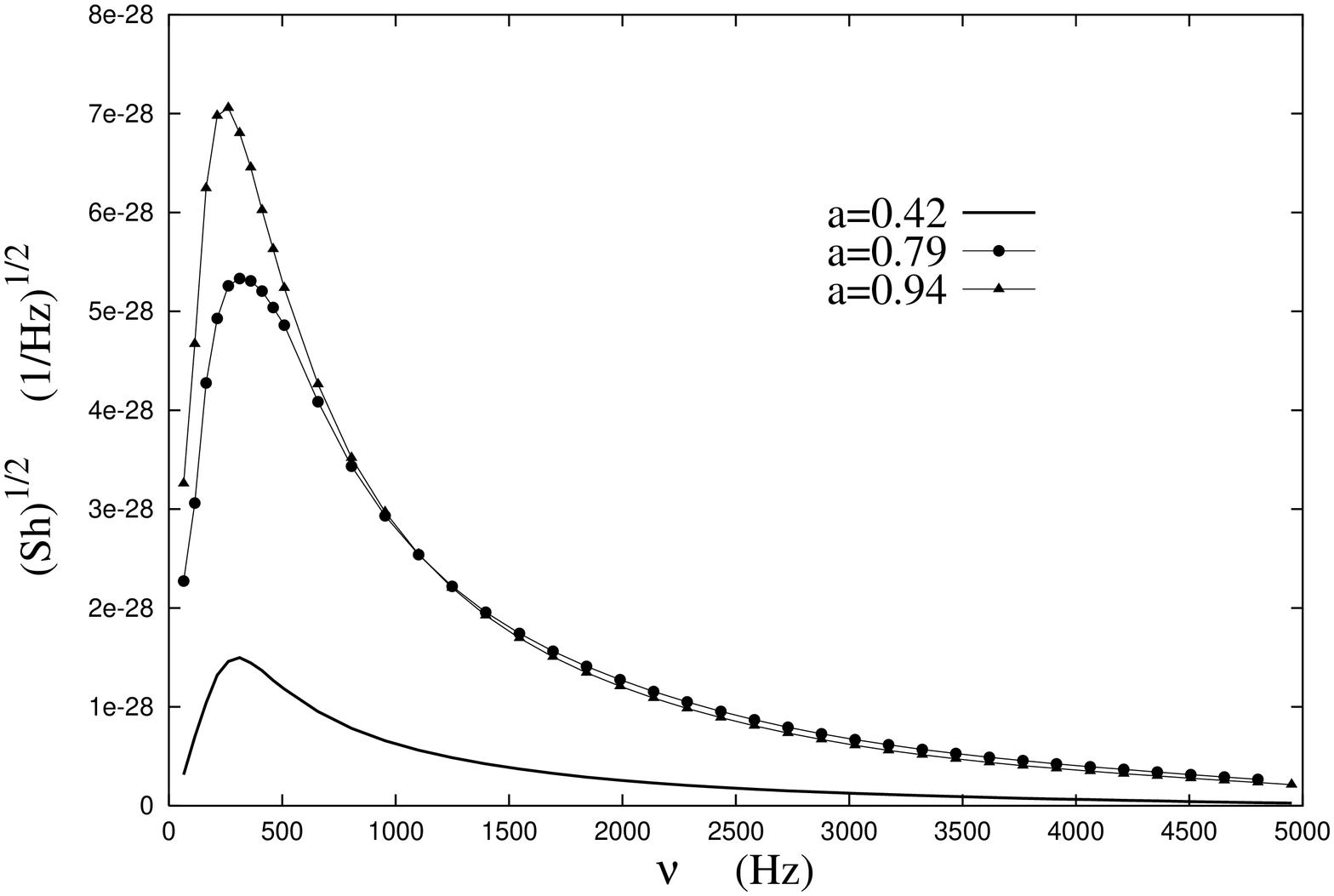,angle=270,width=8cm}}
\vskip 24pt
\caption{\it The closure  energy density \op\Omega_{GW}\cl
(left) and the spectral strain amplitude \op S_h^{1/2}\cl
(right) generated by a cosmological
population of core collapse supernovae leaving behind a black hole,
is plotted  as a function of the observational frequency,
for three values of the angular parameter $a$
and  for a flat cosmology with vanishing cosmological constant.  }
\end{center}
\label{omegash}
\end{figure}

From  eq. (\ref{spec}) we can further derive
the  closure  energy density of gravitational waves
per logarithmic unit frequency,
\be
\label{defome}
\Omega_{GW}=\frac{1}{\rho_c}\cdot\frac{d \rho_{GW}}{d\log\nu}=
\frac{\nu}{c^3 \rho_c}\cdot\frac{d E_{GW}}{dt dS d\nu}
\ee
where \op\rho_c=\frac{3H_0^2}{8\pi G}
\sim 1.9\times 10^{-29}~h^2~g~cm^{-3},\cl
and the spectral strain amplitude 
\be
\label{defstrain}
\sqrt{S_h(\nu)}=
\Biggl[
\Bigl(\frac{2G}{\pi c^3\nu^2}\Bigr)\cdot \frac{d E_{GW}}{dt dS d\nu}
\Biggr]^{1/2}=
\Biggl[
\Bigl(\frac{3~H_0^2}{4\pi^2\nu^3}\Bigr)\cdot\Omega_{GW}
\Biggr]^{1/2},
\ee
which is the quantity to  be compared with the  detectors sensitivity.
The spectral properties of the background produced by
a cosmological populations of core-collapse supernovae leaving behind
a black hole, computed by this procedure,
are shown in figure  \ref{omegash}.
We plot the
closure  energy density  and the strain amplitude
for three selected values of the angular parameter of the formed black
holes, assumed to rotate all at the same speed
since we do not know the distribution of black holes angular momenta.

The  strain amplitude 
exhibits a sharp peak at a frequency
which ranges within $\sim~ [230 - 340]~Hz,\cl depending on  the
chosen value of the angular momentum,
and with a maximum amplitude  
\op
\sqrt{S_{h~max}}\sim
 [1-8] ~\times 10^{-28}~Hz^{-1/2}.\cl
This peak is reminiscent of the excitation of the 
quasi-normal modes  of the formed balck holes, showing that 
the spectral properties of this background keeps memory of the generating
process.

It should be mentioned that since according to eq. (\ref{ratetot})
the event rate is of the order of
\op 3-5~events/s,\cl and since a signal emitted in a collapse to a
black hole has a very short durations 
(cfr. eq. \ref{omegatau}), typically of the order of a few milliseconds,
the background generated by newborn black holes has a shot noise 
character.

The procedure to determine the characteristics  of a stochastic
background described  in this  section can be applied to any population of
astrophysical sources  for which a model for the energy spectrum
is available. For example, in ref. \cite{FMS99b}
we have applied this method to study the background 
produced by  a cosmological
population of young, rapidly rotating neutron stars
that, due to the r-modes instability, are expected
to radiate a large fraction of their rotational energy in gravitational
waves.
A  gravitational  background  can be detected by  cross-correlating the
output of two gravitational antennas for a sufficiently long time interval
(typically one year), and  in refs. \cite{FMS99a} and
\cite{FMS99b} we have computed how interferometric and resonant
detectors operating in coincidence might respond to our predicted
astrophysical backgrounds. 
We find that whereas the sensitivity of the first generation 
of interferometric antennas like VIRGO and LIGO
will be too low to detect these
signals, the planned sensitivity of the advanced version of these experiments
would allow the detection, provided two interferometers with similar
characteristics are located nearby.

\section{Gravitational waves from stars: a perturbative
approach}
A number of numerical simulations of the 
gravitational collapse to a neutron star have shown that,
unlike the case of collapses to a black hole,
the gravitational signals computed  for these processes
strongly depend on the initial conditions, on the equation of state of the
collapsing star,  and on the  details of the collapse
\cite{finn}-\cite{muller}; for this reason,
a model of the energy spectrum emitted in a collapse to a neutron star
is still not available. 
The situation changes if we consider processes occurring after a neutron
star has formed, and we shall now show how some interesting information can
be derived by using a perturbative approach.

The theory of stellar perturbations 
was  formulated in the framework of General Relativity
by Thorne and his collaborators since 1967
\cite{thorne1}-\cite{thorne6}, 
and  it was  successfully applied to determine the
frequencies of the quasi-normal modes of oscillation 
for a wide range of stellar models \cite{lindet,culin}.
Much recently,  the theory   has been
reformulated in close analogy to the theory of black hole perturbations 
\cite{cf1}-\cite{cf7}, and new phenomena have emerged, which 
do not have a newtonian counterpart.
A detailed discussion of the many interesting aspects of this theory is 
beyond the scope of this paper,
where we want to focus essentially on the
characteristics of the gravitational signals emitted in astrophysical
processes.
Thus, we shall rather consider  an application of the theory to
a specific process, 
and show that the emitted signals  exhibit a clear signature
of the nature of the source.
In particular, we shall consider a mass \op m_0\cl which, 
interacting with the gravitational field of a
star of mass \op M\cl deviates from its original trajectory 
moving toward the star, reaches a periastron (the {\it turning point})
and then  moves away in an open orbit.
Under the assumption that \op m_0 << M,\cl
the problem can be solved by using a perturbative approach.
The matching of the solutions of the perturbed equations inside and
outside  the star constitutes the most delicate technical point 
\cite{fergual}. 
Indeed, while  the interior solution is  found by integrating 
the equations for the perturbed metric tensor coupled with the equations of
hydrodynamics,  the exterior solution cannot be found by using
the same tensorial approach because the source term of the equations
for the scattered mass diverges at  the turning point.
However this difficulty can be overcome by introducing 
a suitable wavefunction, related to the Weyl scalar \op\Psi_4,\cl
which carries information on the radiative part of the field.
In ref.  \cite{ferbogual}
the equations describing  the perturbations of a star induced 
by a scattered mass,
have been integrated for a star with a polytropic equation of state
\op p=K\rho^{n},\cl with \op K=100~km,\cl 
\op n=2,\cl  and  central density 
\op \rho_c=3\cdot 10^{15}~g/cm^3.\cl The
radius and mass of the star are, respectively, \op R=8.86~km,\cl and
\op M=1.266~M_\odot,\cl with a ratio \op R/M=4.7\cl
\begin{figure}
\centerline{\mbox{
\psfig{figure=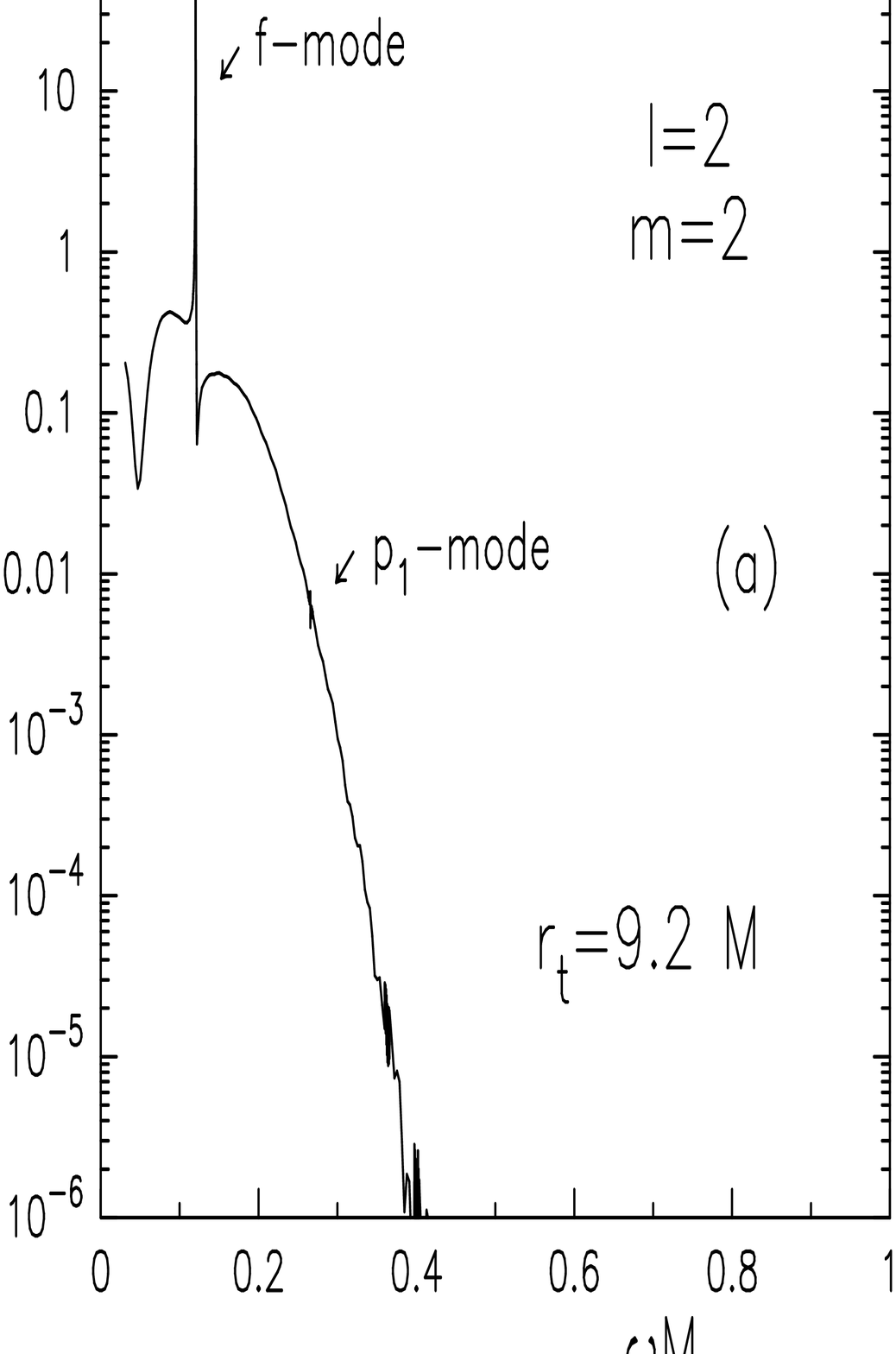,width=10cm,height=6.5cm}}}
\centerline{}
\centerline{}
\centerline{}
\centerline{}
\centerline{}
\centerline{\mbox{
\psfig{figure=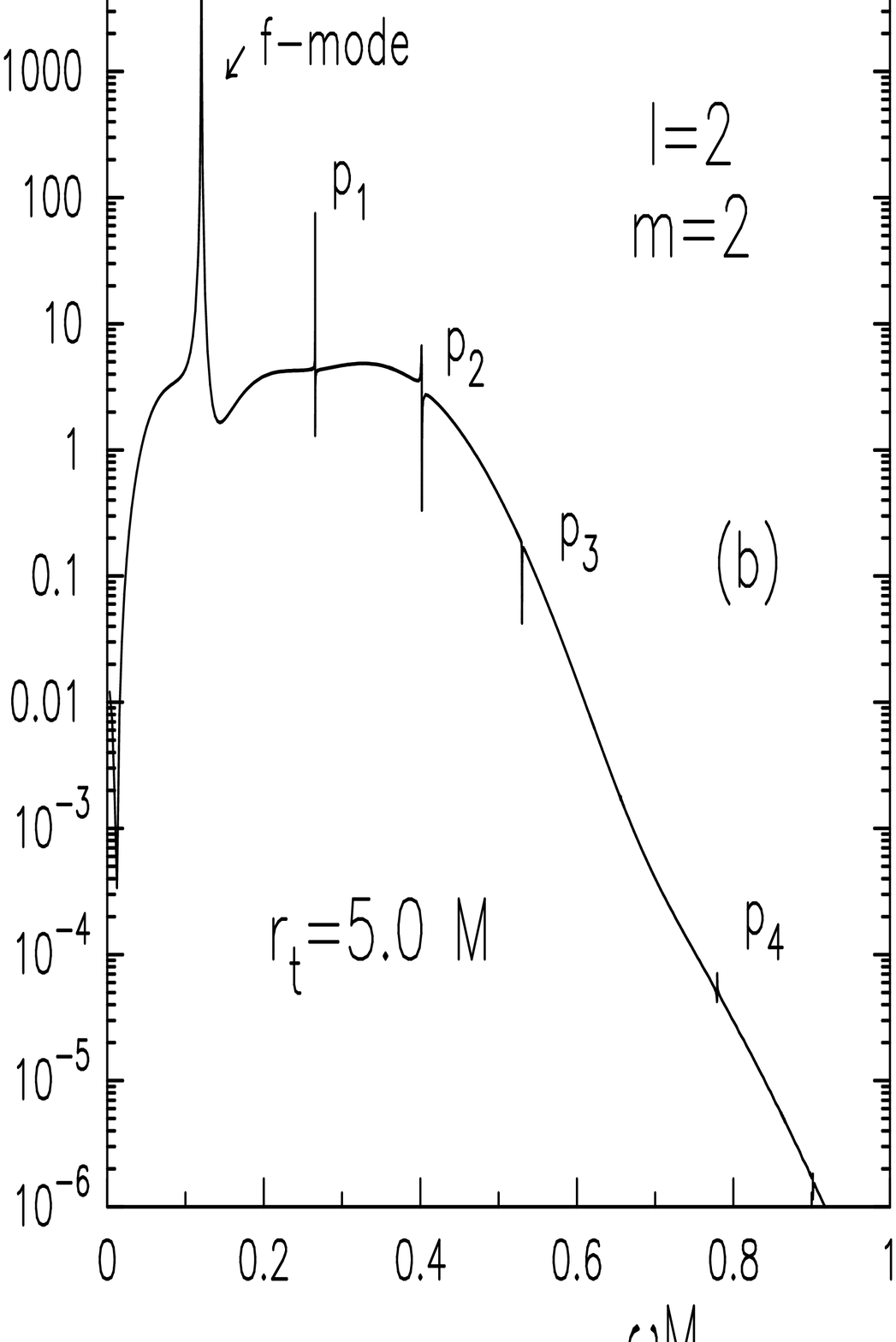,width=10cm,height=6.5cm}}}
\vskip 14pt
\caption{
The \op\ell=m=2$ - component of the
energy spectrum of the gravitational radiation
emitted  when a mass \op m_0\cl 
is scattered by a polytropic star of mass \op M,$
is plotted versus the normalized frequency \op \omega M\cl.
The orbital parameters \op L\cl and \op E\cl are chosen in such a way 
that the turning point is at \op r_t=9.2~M\cl (a) 
and \op r_t=5~M\cl
(b) 
}
\label{energyplot}
\end{figure}
The energy spectrum emitted in gravitational waves
has been  computed for the following sets of orbital paramenters
\begin{itemize}
\item{} a) ~~
$L=5\cl  and \op E=1.007,\cl which correspond to a
turning point located at \op r_t=9.2~M\cl
\item{} b) ~~
$L=5\cl  and  \op E=1.097,\cl so
that the mass can   get closer to the star and \op r_t=5.0~M,\cl
\end{itemize}
and it is plotted in  figure \ref{energyplot} 
versus the normalized frequency \op \omega ~M.$
\footnote{
In figure \ref{energyplot} we plot only the \op\ell=m=2\cl component of the
energy spectrum, because it provides the dominant part of the radiated
energy.
}
$L=L_z/M\cl is the angular momentum of the scattered mass
\op m_0,\cl normalized to $M$, and  \op E\cl is  its energy 
per unit mass.

Figure \ref{energyplot}  shows that
the energy spectra exhibit well defined peaks located at some 
particular frequencies, and in order to understand the physical 
processes that underlie this structure we need to mention 
what are the characteristic frequencies at which a star can oscillate 
and possibly emit gravitational waves.

In newtonian theory, the classification of the modes of oscillations
of a star is based on the behaviour of the perturbed fluid. 
The hydrodynamical equations
show that when a star is perturbed each element of fluid
moves under the competing action of two
restoring forces, one due to the eulerian change in the density
$\delta\rho$, the other due to a change in pressure
$\delta p$.
The modes are classified accordingly:
\g modes,  if the prevailing driving force is due to
$\delta\rho$, \p modes, if it is due to  $\delta p$.
The two classes of modes occupy well defined regions of the
spectrum, and they are separated by the fundamental mode,  the \f mode,
characterized by having  an
eigenfunction that has no nodes inside the star.
In General Relativity stellar oscillations are damped by the
emission of gravitational waves, and 
in addition to the fluid modes there exist modes of
the radiative field.  Indeed, 
it can be shown that fluid motion is either negligible or 
totally absent at the corresponding frequencies.
They are named \w modes,
characterized by high frequencies
and short damping times \cite{kokkoshutz},
and  \s modes, that are slowly damped,  do not excite 
any motion in the fluid 
and are characteristics of ultracompact stars ($R/M \lappreq 2.6$)
\cite{axial}.

Thus,  stars are characterized by a very rich set of possible modes 
of vibration, and it is interesting to ask
whether these modes can be excited in real astrophysical
processes, and  how much energy they carry.
For example, for the model of star considered in figure \ref{energyplot}
the frequency of the fundamental mode
is \op \omega~M =0.120,\cl  the first \p-modes is
\op \omega~M =0.266\cl
and  the first w-mode is \op \omega~M =0.53.\cl
The energy spectra of  figure \ref{energyplot} exhibit 
a pronounced peak at  the frequency of the \f mode,
and more peaks appear, corresponding to
the excitation of the first \p-modes,
if the scattered mass is allowed to
get closer.

It  is interesting to compare this behaviour with that 
of a black hole perturbed by a small mass in a similar scattering process
(see \cite{nakaooharakoji} for an extensive review).
It turns out that the black hole is rather insensitive to these
processes: the energy  is emitted essentially by the scattered mass
as a synchrotron radiation,
and most of it is radiated when the mass
transits through the turning point. 
Indeed, the energy spectrum is peaked at a
frequency which is related to the angular velocity of the mass 
at the turning point as follows
\be
\omega_{r_t}=\ell\Bigl( \frac{d\varphi}{dt}\Bigr)_{r=r_t}.
\ee
In our case a) ($r_t=9.2 M$), the frequency corresponding
to the angular velocity of the mass at the periastron for \op\ell=2\cl
is \op \omega_{r_t}M=0.092.\cl 
The spectrum shown in figure \ref{energyplot}a
has a  peak at  that frequency,
showing that part of the energy is still emitted by the mass as
a synchrotron radiation,  but the
the  very sharp peak which occurs at the
frequency of the fundamental mode is dominant.  
A similar situation arises when
the mass gets closer to the star (case b, $r_t=5 M$, $\omega_{r_t}M=0.221$),
but again the emission is strongly  dominated by the
excitation of the  modes of the star.
Thus, we can conclude that, unlike  black holes,
whose quasi-normal modes  are not excited in scattering processes,
a mass moving around a star in an open orbit can excite
the fluid modes of the star,
to an extent that depends on how close  is the encounter.
The w-modes do not appear to be significantly 
excited in these processes, and
since the contribution of the fundamental mode appears to be the dominant
one,   the emitted gravitational signal is  a
pure note corresponding to that frequency.

\section{ Concluding Remarks}

The study of the gravitational radiation emitted by 
astrophysical sources is an open field to explore, and it requires
the cooperative effort of scientists working in different
fields.   Complex phenomena like the gravitational collapse
of a massive body or the coalescence of binary systems are 
paradigmatic:  powerful computers and specific expertise in
numerical techniques to deal with the
strong field regimes typical of these processes are needed,
as well as
experience in the  physics of phenomena occurring at supernuclear densities
and a deep mathematical knowledge of the equations of gravity.
In addition, since  to detect a signal it is important to know
not only the waveform and the energy it carries,  
but also the rate of occurrence,
the knowledge of the sources distribution and evolution
throughout the Universe, either theoretical 
or based on observations, is also crucial.

Although most of the interesting phenomena that are associated
to the emission of gravitational radiation are highly non linear, 
much can be learnt by the use of approximation schemes. 
For instance  the perturbative
approach  proves extremely useful in providing a physical
understanding of many  processes;  it  helps to interpret and confirm
the results of fully  non linear  simulations,  and
to identify physical mechanisms operating in different
regimes. An example of this synergy of the approaches
is given in section 2, where the peaks of the  energy spectrum
emitted  in a collapse to a black hole, computed by a
fully relativistic numerical simulation,
have been interpreted in terms of
the excitation of the quasi normal modes  predicted by
the theory of black hole perturbations.

Furthermore, we  have shown how theory and observations can be matched 
together to evaluate the spectral properties of the background 
of gravitational waves produced by these sources:
the energy spectrum of each single event
derived by  numerical simulations has been convoluted
with  the star formation rate history  deduced from 
astronomical observations.

The work presented in this paper can be extended in several directions.
We plan to apply the procedure developed to evaluate the 
background of gravitational waves to other cosmological
populations of  astrophysical 
sources, like binary systems formed by neutron stars, black holes and 
white dwarfs \cite{newpaper}.
Furthermore, we shall compute energy and waveforms of the signals  
emitted by stars perturbed by masses orbiting in closed orbits, and
investigate the relation between the amplitude of the emitted
wave and the  equation of state prevailing in the star interior, 
for different star models ranging from sun-like
stars to  white dwarfs and  neutron stars.
Finally, we plan to investigate the role that the excitation of the
quasi-normal modes can play in the coalescence of  binary systems,
and extend all these calculations to rotating stars.


\end{document}